# Why Geosciences and Exoplanetary Sciences Need Each Other

Oliver Shorttle[1], Natalie Hinkel[2] and Cayman Unterborn[2]

1 Department of Earth Sciences & Institute of Astronomy, University of Cambridge, UK

2 Southwest Research Institute, 6225 Culebra Rd, San Antonio, Texas 78228-0510

## Abstract

**The study of planets outside our solar system may lead to major advances in our understanding of the Earth, and provide insight into the universal set of rules by which planets form and evolve**. To achieve these goals requires applying geoscience's wealth of Earth observations to fill in the blanks left by the necessarily minimalist exoplanetary observations. In turn, Earth's many one-offs, e.g., plate tectonics, surface liquid water, a large moon, and life -- which have long presented chicken and egg type conundrums for geoscientists -- may find resolution in the study of exoplanets possessing only a subset of these phenomena.

## Why take the geosciences beyond the solar system

Fields within geosciences are united by study of one planet, the Earth, and its interconnected systems, from the core to the outer edge of its atmosphere. Across these systems we have accumulated decades of data, at the nanometer-scale: through investigations of Earth's material components, and at the planet scale: in recording the oscillations triggered by earthquakes. These data have provided us a deep insight into the forces that shape our planet, and how they have conspired to maintain its habitability over four billion years.

The depth of understanding the geosciences have gained of the Earth is in marked contrast to exoplanetary science, where thousands of planets have been discovered beyond the solar system, but many are only characterised at the level of having a measured mass or a radius (or more infrequently both). Even under the most ideal circumstances, we will have limited information about the planet, extending to knowledge of its orbital parameters and the composition of any atmosphere. Ironically, it may be in death that we learn most about a planet's composition, as it tumbles onto the hot dense remnant of its host star (Xu & Bonsor *This issue*). To understand whether exoplanets are at all like the Earth, insights from geosciences will thus be critical to interpolate between, and extrapolate from, the limited observational data to construct a broader picture of conditions on planets outside the solar system.

As well as fulfilling our curiosity about what alien worlds might be like, an informed reconstruction of their environments is key to ensure we make judicious use of valuable space and ground-based telescope time: which planets should receive the limited hours of observation in an attempt to measure their atmosphere for signs of biological influence?

Only by building links between geoscientists, astronomers, and prebiotic chemists now, will we have the tools to answer this question when the need arises.

Perhaps most excitingly, the current exoplanetary dataset already presents us with a wealth of planetary diversity not seen in our solar System: planets with masses smaller than Mars to those intermediate in mass between Earth and Neptune; with permanent day and night sides; and with surface temperatures of thousands of Kelvin, possibly maintained as permanent magma ocean worlds. These sub-Mars and super-Earth (or mini-Neptune) planets outline a wide range of pressure and temperature conditions unlike any experienced on the planets of our solar System. The bulk compositions of these rocky exoplanets too are thought to span a wide range, many of which are far from Earth-like. Perhaps the best example of this are the so-called "Super-Mercuries", with radii greater than the Earth but with the central Fe-core making up nearly 50% of the planet's total mass (compared to Earth's 33%). With no direct analogues in our solar System, progress here requires new experiments and derivative models that can describe everything from core formation to the evolution of the crust, atmosphere and oceans, to understand the geodynamics and climates of these planetary types new to the geosciences.

For now though, we can carefully extend the models developed to explain Earth's history, interior, surface, and biological dynamics to the broader context of exoplanets, recognising the fundamental aspects of our discipline. One such key Earth system that has been extended to exoplanets is the silicate weathering cycle, which underlies definitions of the habitable zone (Kasting et al., 1993). Most excitingly, on subsequent characterisation of exoplanets we may discover them to deviate from our habitable zone predictions: perhaps they show temperate climates closer to or further from their stars than we thought possible; perhaps temperate climates are found to be vanishingly rare, or most planets enveloped by thick water oceans. Whether our predictions are proven right or wrong, exoplanets will provide a powerful test of our models for Earth evolution. Ultimately, exoplanet characterisation may distinguish those universal properties of rocky planet evolution from those that are highly contingent on chance initial conditions, and so -- for the first time -- place Earth and the life on it into context with its galactic environment.

## Finding what makes Earth Earth-like

Perhaps the most critical question in our search for Earth-like planets is to define what we mean by "Earth-like". This requires a balance between an impossibly specific set of requirements that can never be fully measured, or a superficially general set that may lead to false positives. Of course, all the properties we take to define "Earth-like" derive from only one planet. So which of our planet's characteristics tipped the scale? Our nearby solar system neighbours, Mercury, Mars, and Venus, are clearly non-Earth-like, but also different in enough respects (size, composition, location, biosphere, size/number of satellites, thermal history) that identifying a single cause for these planets' divergent geodynamic and climate states is challenging (see Ehlman et al. 2016). For example, was it the chance input of $^{26}$Al as a heat source to the early solar system that ensured Earth formed with dry land and was not enveloped by thick oceans (Lichtenburg et al. 2019)? Is Earth's temperate climate the result of regulation by life (Höning 2020), or plate tectonics (Foley & Smye, 2018)?

The rather severe limitation of only having one Earth to study is partly mitigated by geological time, over which Earth has been many planets: a magma ocean world, an anoxic greenhouse, snowball, and latterly a planet with an oxygen rich atmosphere and terrestrial and marine biospheres. Information on these past planetary states is only accessible through the rock record, which despite careful study has left many gaps in our understanding of how different parts of the Earth system operated. Ultimately, we do not know to what extent Earth's remarkable present is contingent upon its past: finding possible forks in the evolutionary road should be a key goal of experimentalists and modelers.

While we may only observe an exoplanet at a single snapshot in time, they span a wide variety of ages, with the best characterized rocky planets between 1.4 to 11 Gyr old. This age range may allow us to peek into rocky planet evolution through time, and identify any common trends in terms of climate and geodynamic evolution. By better understanding these exoplanets, we can potentially better infer aspects of the Earth's past and perhaps even its future.

Exoplanets may finally allow us to understand what "Earth-like" means. Planets of different masses and radii, with or without moons, plate tectonics, life, and nitrogen-rich atmospheres, orbiting at different locations, around stars of different masses and composition, with different planetary companions, and in systems of different ages represent nature's parameter sweep of possible conditions for rocky worlds. By characterizing these planets and their host stars over the coming decades, and linking each planet's atmospheric properties to geological processes, it will be possible to discover how robust "Earth-like" is to perturbation of conditions away from those our planet enjoys in our solar system. In turn, we will be able to use this information to better understand the key events and dynamics that have shaped our own planet's history.

## The present and future of exoplanet discovery

There are thousands of known exoplanets, with more being discovered every day. The exoplanet community is currently building state-of-the-art telescopes and instruments in order to propel the next generation towards a better understanding of planetary formation and evolution, as well as the detection of clear biosignatures. Currently in orbit is NASA's Transiting Exoplanet Survey Satellite (TESS, launched 2018), which is providing a precise photometric survey of bright stars across this sky. The CHaracterising ExOPlanets Satellite (CHEOPS, launched 2019) out of the European Space Agency (ESA) is the companion mission to TESS, primarily focused on precisely measuring the radii of known transiting planets, supplemented by ground-based mass measurements. NASA's James Webb Space Telescope (JWST) is slated to launch late 2021 in order to characterize exoplanet atmospheres with a variety of mid- to low-resolution infrared spectroscopy instruments able to measure atmospheric transmission, emission, and phase-curves. In addition, JWST will directly image planets and debris disks and spectroscopically measure debris disks around white dwarf stars. The Nancy Grace Roman Space Telescope out of NASA is set to launch in 2025 and will focus on determining the occurrence rates of planets beyond 1 AU via microlensing, while simultaneously detecting small, transiting planets with short-periods (periods < 2 month). In conjunction with a Coronagraphic Instrument (CGI), directly imaging giant planets and debris disks will be possible, allowing us to better understand how planets form and inherit their final composition. ESA's PLAnetary Transits and Oscillations of stars (PLATO) mission will launch in 2026 and is combining precise photometric detections of transiting planets, particularly rocky planets around mostly solar-like stars, with high-resolution ground-based spectroscopy. The ESA Atmospheric Remote-sensing Infrared exoplanet Large-survey (ARIEL) has teamed up with the NASA Contribution to ARIEL Spectroscopy of Exoplanets (CASE) mission to launch in 2028, in order to characterize ~1000 planets, their atmospheres, and atmospheric thermal properties in order to determine planetary formation, migration, and evolutionary histories.

All of these exoplanet missions, as well as those still on the horizon for the years 2040-60, will be measuring properties that are crucial to characterizing rocky and gaseous exoplanets. Figure 1 shows how stellar and exoplanet observations feed into the myriad geoscience disciplines, demonstrating the many possible synergies that will expand our understanding and characterization of rocky exoplanets. We note that there are a number of key planetary observables, such as atmospheric-interior exchanges, reflectance spectra, and planetary interior temporal variability, which currently have no planned exoplanetary mission or ground-based input. This lack of upcoming data in these areas outlines the strong need within exoplanetary science for additional missions that can fill these niches, with mission teams composed of observers and modelers to quantify the observational consequences of these not well understood processes. Overall, Figure 1 provides an interdisciplinary map between the geosciences, exoplanet science, and astronomy. It emphasises how the connection between observables and models is crucial for holistic exoplanetary characterization.

Article 1 in the "Geoscience Beyond the Solar System" issue of Elements magazine, v17 No4

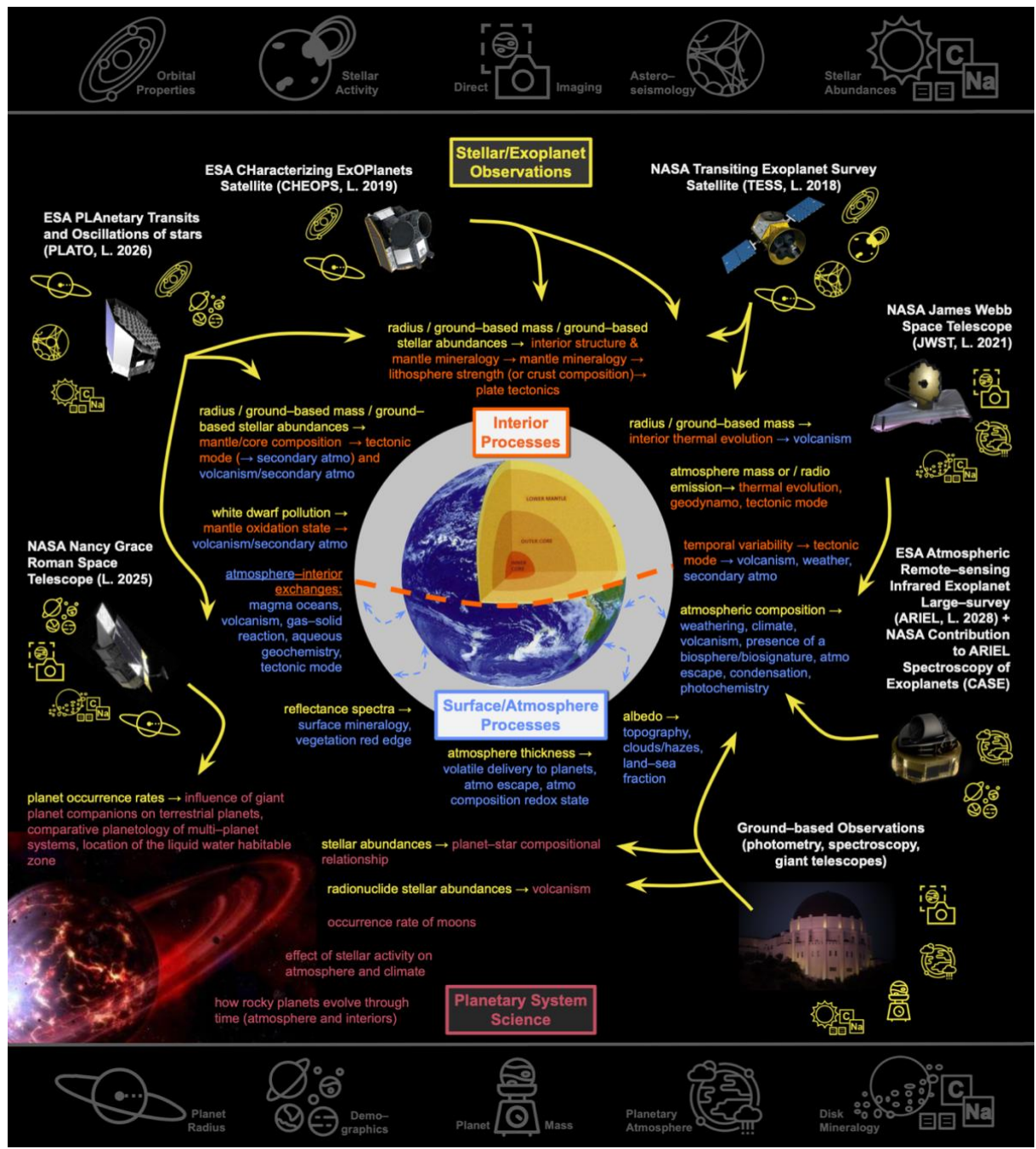

Figure 1 caption: We have put an image of a planet and atmosphere at the center -- the planet is bisected to show the planet's cross-section and interior processes (orange) at the top while the surface and atmospheric processes (blue) are at the bottom. On the bottom left is a neighboring planet, to represent planetary system science and exoplanet demographics (red). The outer edge of the figure shows NASA and ESA missions with their respective launch dates (denoted as "L."), as well as ground-based observations. Surrounding the missions are icons (yellow) depicting the properties they have or are expected to measure, where the icons are defined at the top and bottom of the figure. The missions have large arrows connecting their specific observables to either 1) key interior processes, which may then contribute (via smaller arrow flow downs) to our understanding of surface and/or atmospheric processes, or 2) key surface and/or atmospheric processes. Some icons were used via the Noun Project, specifically attributed to Pascal Heß, Lufti Gani Al Achmad, Eucalyp, Template, Alison, Andreis Kirm, and Yarden Gilboa.

## Geosciences beyond the solar system

As the list of missions outlined in Figure 1 shows, exoplanetary science is moving on from discovery of exoplanets to the characterisation of their formation, composition, and evolution. This marks a rapid change within the discipline, one which requires the investment of geoscientists to help in interpreting data on planetary mass, radii, atmospheric compositions, and more in terms of the geological histories of these planets. Geoscience has been spoilt by the rich data Earth provides of its surface and interior and the models developed to explain such observations are consequently tuned to match this detail with high-precision. Applying these models to exoplanets will not always come easily, when far fewer of the governing parameters are known and we are asked to match observations with much greater uncertainty than we have for the Earth. However, meeting this challenge represents an excellent opportunity to investigate whether the underlying physics, chemistry (and biology) of our Earth system models capture fundamental planetary processes, and what parts of them are Ptolemaic in construction.

Exoplanetary science also emphasises the need for new experimental work on planetary materials. Experiments on partial melting and phase equilibria have barely reached core-mantle boundary conditions to understand the metal-silicate partitioning of elements. However, rarely has this been achieved for systems outside the pressure, temperature, and compositional range found on the Earth. Yet, we are now faced with the need to understand core formation on planets twice as massive as Earth. How would core formation or mantle convection or partial melting operate under such extreme conditions? Only new experiments, whether laboratory based or theoretical, can answer such questions, and in doing so will provide valuable information for constraining Earth's behaviour as well.

An excellent example of the relationship between exoplanetary sciences and experimental studies are the recent measurements of the UV cross section of water at habitable temperatures (Ranjan et al. 2020). The work, motivated by the need to understand atmospheric chemistry on abiotic planets, provides new data that can also be used to model the early Earth. As the clamour from exoplanetary sciences grows for constraints on material properties at conditions relevant to the diversity of rocky planets in the galaxy, the geosciences will have many more opportunities to contribute experimental expertise.

## This Elements issue

As this issue emphasises, exoplanetary sciences is maturing into a highly-diverse discipline, uniting astrophysicists, geoscientists, planetary scientists, and prebiotic chemists. There is much to be gained for each of these core disciplines by studying exoplanets, but barriers to effective cross-disciplinary collaboration exist. Immediately obvious at any exoplanet meeting is each community's use of jargon. At the end of this chapter we include a glossary of key terms that arise within this thematic issue, so as to limit the barriers to effective communication.

This issue of Elements also provides a primer on how several key fields of geoscience are contributing to exoplanetary science, and outlines future needs. We begin with investigating the compositional diversity of rocky exoplanets, and how the link between stellar composition and planetary composition can allow the prediction of mineralogy and geological outcomes (Putirka et al. *This issue*). The following chapter on polluted white dwarfs shows the remarkable experiment nature has performed, allowing us a glimpse of planetary compositions directly: throwing rocky material onto the remnants of dead stars has provided evidence of differentiated and volatile rich planetesimals having formed around other stars (Xu and Bonsor, *This issue*). We next look at the dynamics of exoplanet interiors, and how their diverse tectonic regimes may map through to surface conditions in a way that may be detectable (Ballmer & Noack, *This issue*). Finding planets with temperate climates is a primary aim of exoplanetary sciences, and the next chapter on climates emphasises how the richness of the parameter space occupied by exoplanets will allow testing of our general understanding of the physics and chemistry governing planetary climate (Komacek et al., *This issue*). The lens through which we will view many geological processes on exoplanets will be their atmosphere, and the next chapter discusses atmospheric gain, loss and evolution and the challenges we will face in making the requisite spectral observations for rocky-planet atmospheres (Schaefer & Parmentier, *This issue*). The last chapter concerns the search for life, and emphasises how the study of prebiotic chemistry and the emergence of life on Earth has now become a field that is linked to the quest for life in the galaxy (Rimmer et al. *This issue*).

## Glossary

*Words in italics are defined elsewhere in the glossary.*

**Abiogenesis**: The formation of the organic molecules leading to the origins of life from non-life, following an origins of life scenario.

**Abiogenesis Zone:** The annulus around a host star inside/outside of which *abiogenesis* cannot occur, given a specific origins of life scenario.

**Abiotic**: Not produced by life.

**Accretion**: The growth of planets through the collision of smaller objects that range is in size from dust particles to *planetesimals* to *planetary embryos*.

**Albedo**: The amount of stellar radiation reflected by a planet. A "white body" has an albedo of 100%, reflecting all incident radiation back into space.

**Aquaplanet**: A planet with a surface entirely covered by a liquid water ocean.

**Biosignature**: An object, substance, and/or pattern that is produced by life and could be identified on an *exoplanet* as non-abiotic in origin.

**biosignature, Agnostic** - A biosignature indicating the presence of life, independent of any specific biochemistry.

**biosignature, Anti-**: A molecule or phenomenon that is detectable on an exoplanet and indicates the absence of life.

**biosignature, A-:** A molecule or phenomenon that is detectable in an exoplanet and does not constrain the presence *or* absence of life.

**Black body (or blackbody):** any object that absorbs all radiation that it receives, reflecting none of it back into space; black bodies have an *albedo* of 0%.

**Calcium-aluminum-rich inclusions (CAIs):** The most *refractory*-rich grains known to exist and also believed to be the first solids formed in the solar system.

**Chondrites:** A class of rocky meteorites that contain distinct round grains called "chondrules"; includes the sub-classes *ordinary-* and *carbonaceous-chondrites.*

**Chondrites, Carbonaceous**: Carbon-rich chondrites having bulk compositions that are similar to the Solar *photosphere* when *volatile* elements (e.g., H, He, N, Ar, Xe) are ignored.

**Chondrites, CV**: The most *refractory*-rich group among the carbonaceous chondrites.

**Chondrites, Ordinary**: The most frequent meteorite occurrence on Earth; more *volatile* poor than the *carbonaceous chondrites*.

**Chondritic**: Any rock that has elemental abundances similar to *chondrite* meteorites.

**Circumstellar**: Material that is orbiting around a central star.

**Cloud Condensation Nuclei (CCN)**: Small (sub-micron sized) particles on which water vapor can condense and form a cloud.

**Convection:** Heat transfer accommodated by the bulk motion of material, usually driven by the temperature dependence of density ("thermal convection"), but also by compositional density anomalies ("thermo-chemical convection").

**Convection, mantle**: Solid-state convection in the rocky shell of a planet at rates of cm/year to m/year. Accommodated by solid-state deformation (creep) of silicate minerals.

**Core**: The Fe-metal-rich interior of a rocky planet and source of its *geodynamo*. On Earth, separated into a solid inner and liquid outer core.

**Coronagraph**: A telescope instrument that blocks the light from a star, enabling the direct observation of nearby objects.

**Crust**: A compositionally distinct layer of a planet forming its surface, produced by *partial melting* of the underlying *mantle* and the eruption of these melts onto the surface as volcanoes, or their freezing inside the crust.

**Dex**: the decimal exponent system, e.g., as used in astronomy to report stellar compositions, more commonly known as "order of magnitude"; a difference of a factor of ten is "1 dex".

**Differentiation**: A process that separates elements; in a planetary body, refers to separation into different phases by their chemical affinity, aided by gravity this leads to formation of metal *cores*, rocky *mantles* and *crusts*.

**Envelope**: The gas or fluid envelope covering a planet, comprising material between its rocky surface and the top of its atmosphere.

**Equilibrium temperature**: The temperature an object has when incoming radiation and outgoing radiation are balanced, assuming it is a perfect black body. For Earth this temperature is 255 K.

**Exo-biosphere**: A biosphere on an *exoplanet*, relevant to discussing potentially observable effects of life on exoplanets.

**Exoplanet**: A planet found outside the Sun's orbit, typically, although not exclusively, orbiting another star.

**ExoEarth**: An Earth-size/mass *exoplanet* that receives a similar amount of incident radiation to Earth and orbits a star similar to the Sun.

**Feedback**: A process that either further amplifies or mutes a perturbation. Positive feedbacks lead to amplification, while negative feedbacks diminish effects.

**Galactic chemical evolution**: The time- and location- dependent composition of the interstellar medium, as stars synthesize and disperse heavier elements through the galaxy.

**General Circulation Model (GCM)**: A numerical model that solves the fluid equations of motion under conditions relevant to planetary atmospheres.

**Geodynamo (planetary dynamo)**: Process that generates the Earth's (planet's) magnetic field, for a rocky planet by convection in its metallic *core*.

**Habitable zone (also water habitable zone)**: Region around a star in which a planet that has Earth-like atmospheric composition and geochemical cycling could have surface conditions permissive of surface liquid water.

**Hot Jupiters**: Gas giant planets with *equilibrium temperatures* between 1000 – 2000 K. Above 2000 K, exoplanets are classed as "Ultra-hot" Jupiters.

**Ice-albedo feedback**: A process by which the formation of ice enhances planetary *albedo*, cooling the surface and leading to enhanced ice formation.

**Hadley cell**: A large-scale circulation that occurs in the tropics of Earth, characterized by upwelling of air near the equator and downwelling in the sub-tropics.

**Instellation**: The amount of stellar radiation that a planet receives from its host star. For planets around the Sun, referred to as "insolation".

**Kuiper-Belt Object**: A group of minor, rocky and icy objects in the outer solar system, at a distance of roughly 20-50 AU from the Sun, of which Pluto is among the largest.

**Lithopanspermia**: proposed mechanism for transfer of life from one planet to another via meteorites, ejected by impacts or volcanism.

**Lithosphere**: A mechanically-distinct layer of a planet, forming its cold and stiff rocky outer shell. At its top it includes the *crust,* and its base may extend down to include the top (non-convecting) part of the planet's *mantle*.

**M dwarf**: The smallest class of star (i.e. with fusion in their core), with masses ranging from 0.08 - 0.6 times that of our Sun.

**Magma ocean**: A region of large-scale silicate melting on a planetary object. Magma oceans may either be sub-surface, as on Jupiter's moon Io, or extend fully to the surface at sufficiently high heat flux.

**Mantle**: A compositionally distinct layer of a planet forming the main rocky envelope below the *crust* and above the metallic *core*.

**Mantle plume**: A buoyant convective upwelling of *mantle* rock, usually driven by its excess heat compared to surrounding *mantle*; the excess heat is thought to derive from the core-mantle boundary and may induce surface volcanism (e.g., Hawaii on Earth; Olympus Mons on Mars).

**Mass-Radius Relationship**: Models that relate the mass to the radius of a planet, via assumptions about their bulk composition and structure (the layering into *envelopes*, *crusts*, *mantles* and *cores*).

**Metals (Astronomy):** Any elements heavier than He.

**Metallicity:** The abundance of *metals (Astronomy)* in an object.

**Mini-Neptunes**: Planets that are transitional in mass between *super Earths* and Neptune, which has 17 times Earth's mass.

**Mobile lid**: A surface tectonic regime of rocky planets, where the *lithosphere* undergoes significant lateral movement, e.g., via plate tectonics on Earth; contrast with a *stagnant lid*.

**Orbital period**: time required for a planet to complete one trip around its central star.

**Oxygen fugacity ($fO_2$)**: A measure of how oxidising a system is. Strictly, it is the partial pressure of $O_2$ in a system, where $O_2$ is a real (rather than ideal) gas. That system could be a rock or a planetary atmosphere. In a planetary *mantle*, $fO_2$ can be derived from the different amounts of Fe existing in its different valence states. $fO_2$ is often reported with respect to a simple reference system, e.g., co-existing fayalite-magnetite-quartz (FMQ).

**Partial melting**: The process whereby a solid assemblage of minerals melts to produce both a liquid and a new solid assemblage, each of different composition to the starting solids.

**Photosphere**: The lowest portion of the star's atmosphere from which visible light is radiated.

**Planetesimal**: A km-sized rocky body formed in the early stages of planetary growth, held together by self gravity.

**Planetary embryo**: A 1000 km (Moon to Mars) sized object that gravitationally dominates its region of the *protoplanetary disk*.

**Plate tectonics:** A theory describing the motion of rigid plates on a sphere, where deformation is localised at plate boundaries. On Earth, this explains the creation of new *crust* and *lithosphere* at mid-ocean ridges and the destruction of *crust* at *subduction* zones.

**Primitive equations**: A simplified version of the fluid dynamical equations relevant to atmospheres that are hydrostatic and where the atmosphere is thin relative to the size of the planet.

**Protoplanetary disk**: A *circumstellar* disk of gas and dust lasting typically < 10 Myr out of which planets form.

**Radial velocity method**: The detection of exoplanets, and estimation of their mass, by the doppler shifts they induce in their host star's spectra during their orbit.

**Refractory**: Elements or phases that require high temperatures to melt, evaporate or sublimate (as opposed to *volatiles*), including elements such as Ca, Al, Ti, and Fe.

**Roche limit**: See *Tidal radius*.

**Runaway greenhouse**: An atmosphere that has undergone a positive *feedback* between greenhouse gas concentration and temperature such that the surface of the planet cannot sufficiently cool, resulting in the loss of surface liquid water and high surface temperatures.

**Secondary eclipse**: When a planet passes behind its host star, blocking its emitted (or reflected) light from reaching Earth. Immediately prior to secondary eclipse the planet's dayside faces the Earth.

**Silicate-weathering feedback**: A geochemical *feedback* in which carbon dioxide is removed from the atmosphere and incorporated into silicate rock, with the reaction's efficiency increasing with temperature and precipitation.

**Solar twins**: Stars nearly identical to the Sun with respect to temperature, surface gravity, and *metallicity*. "Solar analogs" are similar to the Sun, although less strictly so.

**Solar-type stars:** Stars that have hydrogen fusion in their core; F, G and K spectral type stars, are all considered "Solar-type" (see *Stellar class*).

**Spectroscopy**: The study of wavelength-dependent absorption or emission of matter interacting with light, e.g., of a planetary atmosphere or stellar photosphere.

**Stagnant lid**: A surface tectonic regime of rocky planets where, in contrast to *mobile lid* planets, the *lithosphere* forms a single rigid and immobile plate, e.g., Mars.

**Stellar class:** Stars are classified by spectral type according to their temperature (or color) and luminosity. These are presented in a Hertzsprung-Russell Diagram of color vs. absolute magnitude (brightness, corrected for distance). Spectral classes are O, B, A, F, G, K, and M, where O is the hottest. Each class is subdivided further by the numbers 0-9, where 0 is the coolest, and subdivided further by roman numeral I-V where I is the brightest. The Sun is a type G2V star.

**Subduction**: A necessary consequence of *mobile lid planets*; where two *lithospheric* plates collide, mountains are created and the denser plate will sink into the *mantle*.

**Super Earths**: Planets roughly 2 to 10 times Earth's mass, potentially having rocky surfaces.

**Tidal radius (or Roche limit)**: The critical distance where the gradient in gravitational force from an object (e.g., a *white dwarf*) is large enough to rip apart the second object (e.g., a *planetesimal*).

**Tidally-locked**: Where the rotation rate of a planet is locked in a resonance with its orbital period. The value is typically 1:1 ("tidally-synchronized") meaning that one side of the planet permanently faces its star, although ratios of 2:1 and 3:2 (e.g., Mercury) are also considered tidally-locked.

**Transit Method**: The detection of an *exoplanet* by its passage in front of its host star causing a dimming of the stellar light received at Earth. This method is capable of constraining an *exoplanet*'s radius and orbital period.

**Urey Ratio (Ur)**: The ratio of the total internal heat production (e.g., from inner *core* crystallisation and radioactive decay) to total surface heat flux.

**Volatiles**: elements that readily evaporate or preferentially partition into a vapor phase, such as H, He, N and Ar (also called "atmophile elements"); at high temperatures elements such as Na or K are also volatile.

**White dwarf**: The final evolutionary stage for stars with original (main sequence) masses less than 10 solar masses. Transitioning to white dwarfs stars will lose half or more of their original mass; once they become white dwarfs, they are about the size of Earth.

**white dwarf, Polluted**: White dwarfs that show elements heavier than He in their atmospheres